\documentclass[doublecol]{epl2} 

\usepackage{multirow}
\usepackage{amsmath}
\usepackage{amssymb}
\usepackage{extarrows}
\usepackage{bbm}
\usepackage{layouts}
\usepackage{url}

\newcommand\emin{E_{\mathrm{min}}}
\newcommand\epeak{E_{\mathrm{peak}}}
\newcommand\ehigh{E_{\mathrm{high}}}

\newcommand\ecut{E_{\mathrm{cut}}}

\title{Entropy of unimodular Lattice Triangulations}
\shorttitle{Entropy of unimodular Lattice Triangulations} 

\author{Johannes F. Knauf\inst{1} \and Benedikt Kr\"{u}ger\inst{1} \and Klaus Mecke\inst{1}}
\shortauthor{J. F. Knauf, B. Kr\"uger and K. Mecke}

\institute{                    
  \inst{1} Institut f\"{u}r Theoretische Physik, FAU Erlangen-N\"{u}rnberg, Staudtstr. 7, 91058 Erlangen, Germany
}

\abstract{
  Triangulations are  important objects of study in combinatorics, finite element simulations and  quantum gravity, where its entropy is crucial for many physical properties. 
  Due to their inherent complex topological structure even the number of possible triangulations is unknown for large systems.
  We present a novel  algorithm for an approximate enumeration which is based on calculations of the density of states using the Wang-Landau flat histogram sampling.  
  For triangulations on two-dimensional integer lattices we achive excellent agreement with known exact numbers of small triangulations as well as an improvement of analytical calculated asymptotics. The entropy density is $C=2.196(3)$ consistent with rigorous upper and lower bounds.  
  The presented numerical scheme can easily be applied to other counting and optimization problems.
}

\begin{document}

\maketitle

\section {Introduction}

Triangulations of spaces are relevant for a broad range of physical phenomena. 
They serve as discretisation of all kinds of surfaces, hypersurfaces and volumes  \cite{DeLoera_2010}, yielding applications of computational geometry in physics, material science, medical image processing  or even in computer graphics and visualisation \cite{Edelsbrunner_2000,Jansen_2001,Krahnstoever_2004,Kolingerova_2001,Kolingerova_2004}. 
Many physical systems can be described by random surface models \cite{Froehlich_1985} -- based on random triangulations. For instance, biological membranes and vesicles can be modelled using triangulated surfaces with curvature-dependent Hamiltonians \cite{McWhirter_2009,Koibuchi_2003,Koibuchi_2010,Kroll_1992,Gompper_1995,Gompper_1997}.

Triangulations are also used as a random graph model for real world networks: 
Random Apollonian networks \cite{Andrade_2005,Zhou_2005,Song_2012}, which are the dual graphs of classical Apollonian packed granular matter and therewith triangulations, show both small-world and scale-free behaviour, as many real world networks.
The properties of triangulations of closed surfaces with arbitrary genus are of much interest, since each graph can be embedded into a closed surface with high enough genus \cite{Kownacki_2004,Aste_2012}.

The (Causal) Dynamical Triangulation approach even tries to describe quantum gravity from scratch with an ensemble of random space-time triangulations as their central entity \cite {Ambjorn_2005}. 
Based on a discrete version of general relativity, where spacetime is approximated by triangles or higher-dimensional analogues, the curvatures become determined purly by the topological structure of the underlying triangulation, e.g. the number of triangles.
The resulting action of the theory can be used to extract a phase diagram and observables - in a path-integral like sum over histories approach  \cite {Ambjorn_2012}.

For an exact evaluation of such quantum gravity models, an enumeration of all possible triangulations would be necessary. However, efficient enumeration of triangulations is an open problem in combinatorics \cite {Aichholzer_1999,Rambau_2002,Aichholzer_Olympics}. 
There is a comparably efficient enumeration algorithm for the special case of planar lattice triangulations at least for small system sizes \cite{Kaibel_2003}. 
Together with the known upper and lower bounds on the number of lattice triangulations this yields a perfect test case for the evaluation of new approximation methods.

In this work we demonstrate, that the Wang-Landau algorithm \cite{Wang_2001} can also be used for counting lattice triangulations approximately but accurately.
Those flat histogram Monte Carlo methods \cite{Lee_1993,Wang_2001} have gained broad attention in statistical physics during the last years.
As well as other Markov chain Monte Carlo methods they have already been applied also for approximate counting state spaces for problems in physics and informatics \cite {Jerrum_1996,Kenyon_1996,Ermon_2011}.

Using this approximate counting scheme we are able for the first time to calculate for large systems the entropy density of lattice triangulations and compare its scaling with analytical bounds obtained in \cite{Kaibel_2003}.
The presented enumeration scheme can also be applied on other physical problems where the number of states of states with certain properties is important, e.g. calculating the degeneracy of the ground state (and thereby the residual entropy) plays an important role for checking the third law of thermodynamics \cite{Chow_1987,Berg_2007}.

\section {Lattice Triangulations}

\begin{figure}[ht]
  \centering
  \includegraphics[width=0.455\columnwidth]{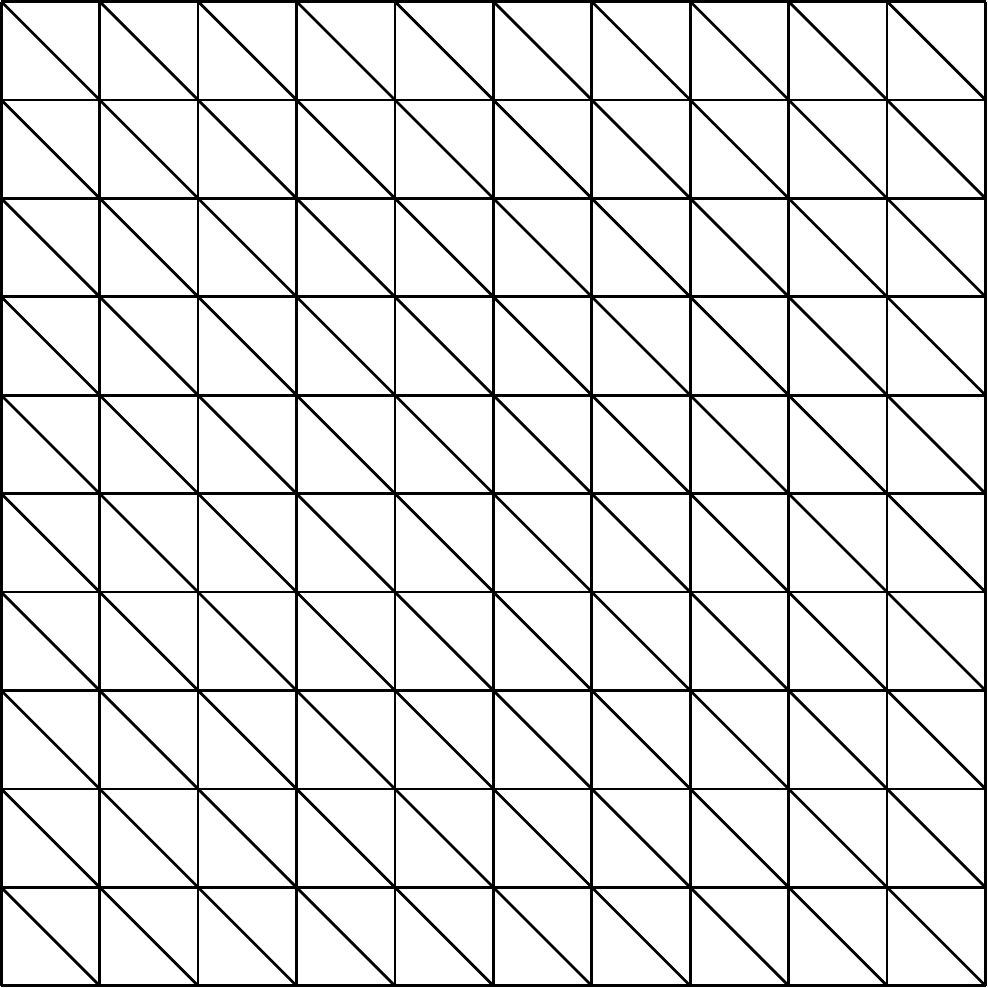}
  \hspace{0.025\columnwidth}
  \includegraphics[width=0.455\columnwidth]{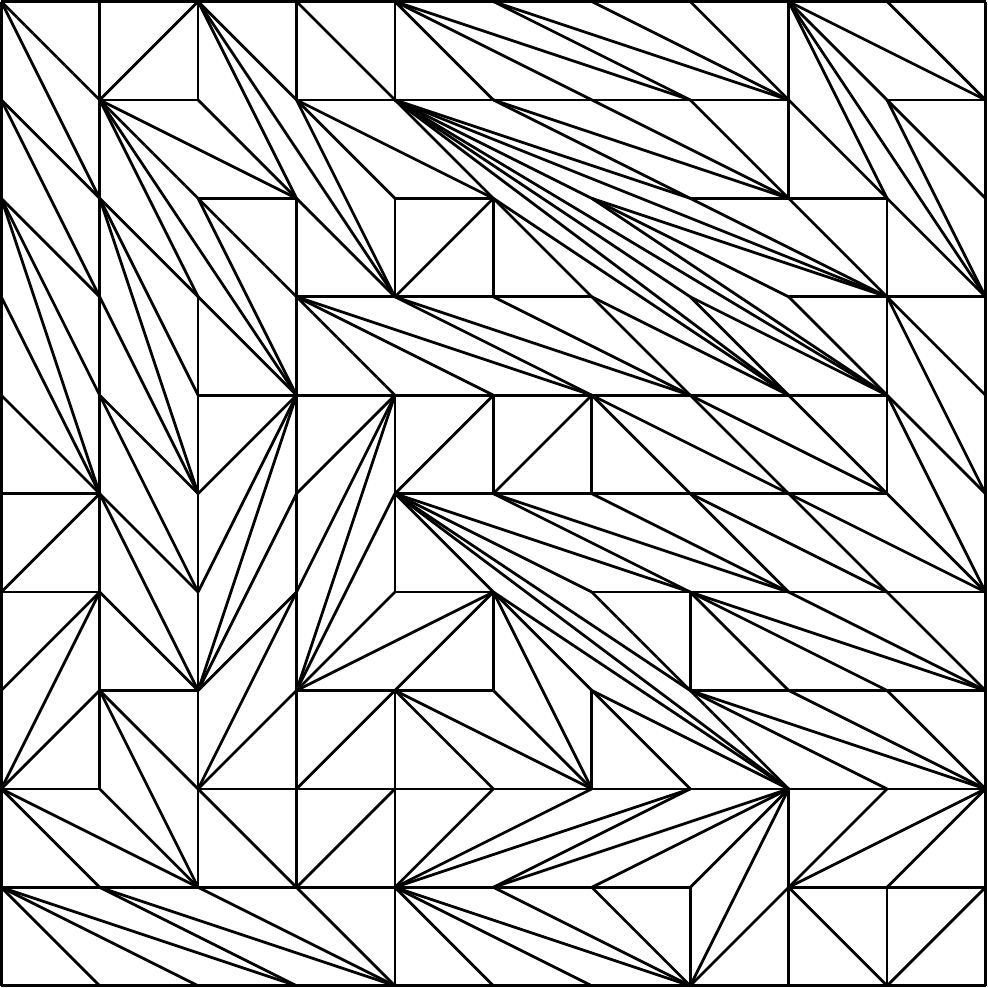}\\
  \vspace{0.025\columnwidth}
  \includegraphics[width=0.455\columnwidth]{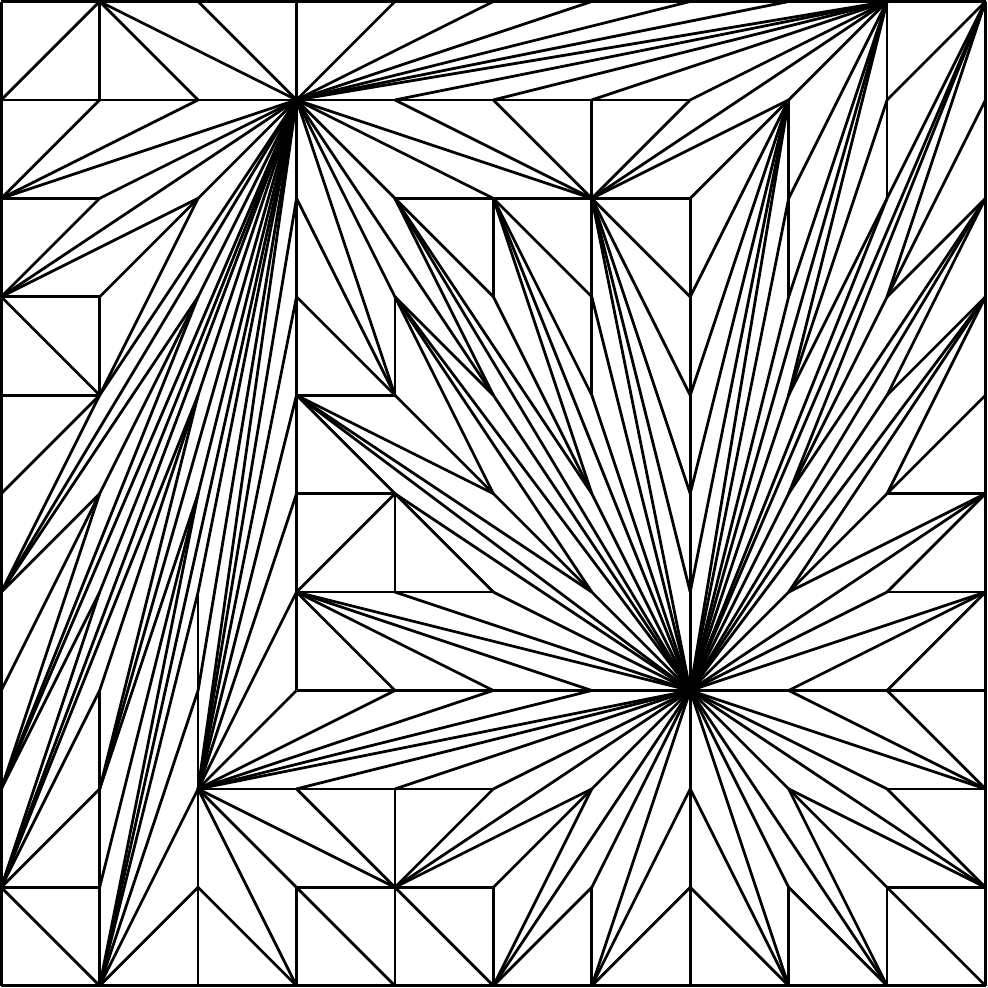}
  \hspace{0.025\columnwidth}
  \includegraphics[width=0.455\columnwidth]{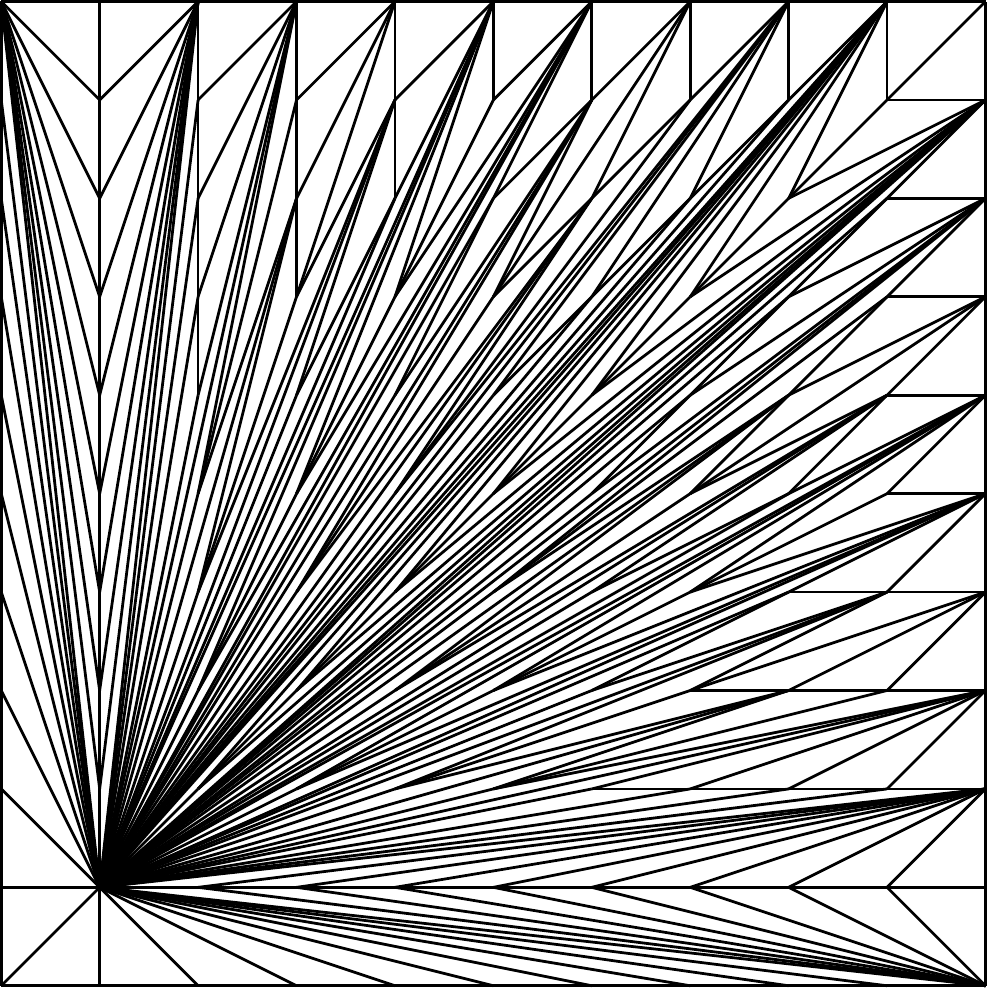}\\
  \caption{Examples for $10\times 10$ lattice triangulations. From top left to bottom right: ordered, regular ground state of the lattice triangulations; random lattice triangulation; immobile lattice triangulation;  lattice triangulation with high vertex degree $E$.}
  \label{fig:triang_examples}
\end{figure}

We follow the definitions given in the book by De Loera, Rambau and Santos \cite{DeLoera_2010}. 
A two-dimensional triangulation is the tessellation of a convex subset of the Euclidean plane into triangular building blocks so that triangles only intersect on their boundaries.
An $m\times n$ lattice triangulation is a full triangulation of a grid $P_{m,n}=\{0,\dots,m\}\times \{0,\dots,n\}$, where full denotes the property that each lattice point is simultaneously a corner of a triangle. 
It is unimodular, i.\,e.\ all triangles have constant area $A=1/2$ \cite{Kaibel_2003,Caputo_2013}. 
The number of vertices is $N_{\mathrm{points}} = (m+1)(n+1)$, the number of triangles is $N_{\mathrm{triangles}} = 2mn$ and the number of edges is $N_{\mathrm{edges}} = 3mn+m+n$ \cite{Kaibel_2003,Caputo_2013}. 
Fig.\,\ref{fig:triang_examples} show examples for $10\times 10$ triangulation.

Edges incident with two triangles creating a convex quadrangle can be flipped into the alternative diagonal of the surrounding quadrangle. 
Each triangulation of a vertex set can be transformed into any other triangulation by a finite number of these diagonal edge flips \cite{Lawson_1972}, hence flips are an ergodic operation on the set of all $m\times n$ triangulations.
In lattice triangulations the surrounding quadrangle of an edge is convex iff it is a parallelogram, which reduces the calculation effort for flippability checks.

\begin{figure}[ht]
  \centering
  \raisebox{0.05625\columnwidth}{\hspace{0.05625\columnwidth}\includegraphics[width=0.1875\columnwidth]{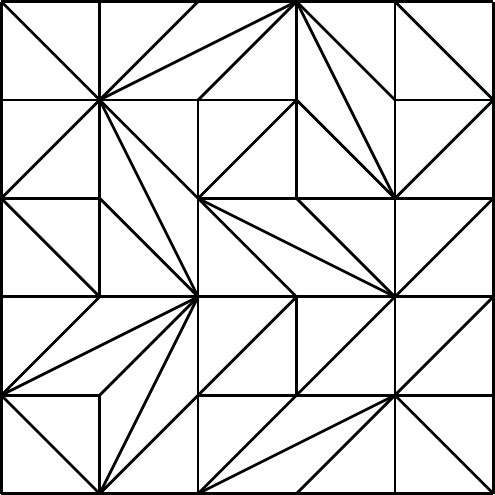}\hspace{0.05625\columnwidth}}
  \hspace{0.025\columnwidth}
  \raisebox{0.0375\columnwidth}{\hspace{0.0375\columnwidth}\includegraphics[width=0.225\columnwidth]{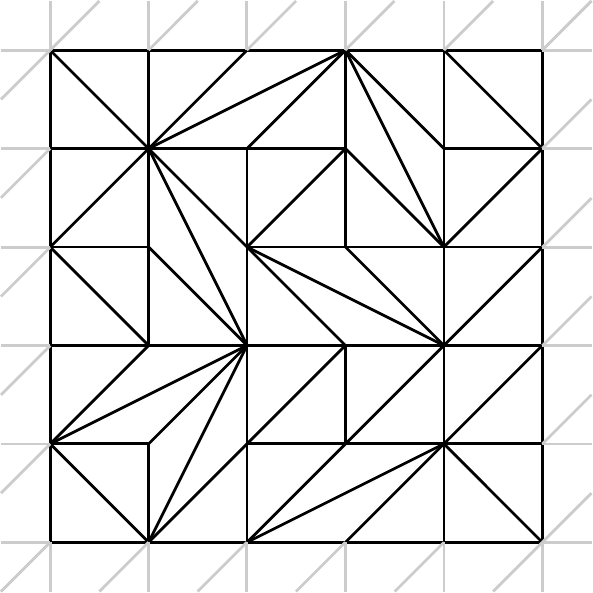}\hspace{0.0375\columnwidth}}
  \hspace{0.025\columnwidth}
  \raisebox{0.0\height}{\includegraphics[width=0.3\columnwidth]{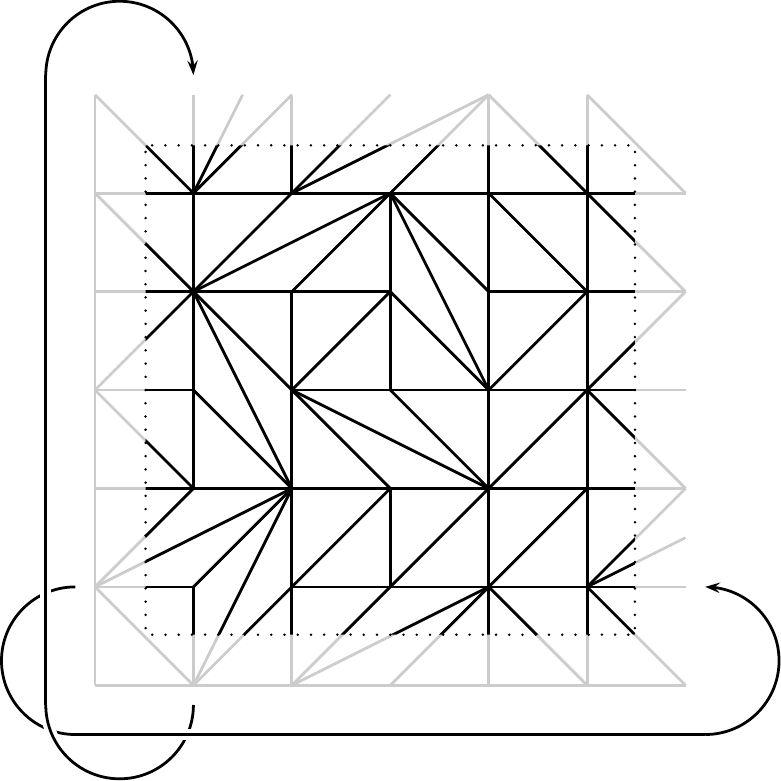}}
  \caption{Different types of boundary conditions for lattice triangulations: open boundary conditions, fixed boundary conditions and periodic boundary conditions.}
  \label{fig:boundaries}
\end{figure}

There are several possible choices of boundary conditions (BC), for instance free, periodic and fixed BC as shown in Fig.\,\ref{fig:boundaries}.
For the latter case the different triangulations are embedded into a bigger lattice equipped with a fixed triangulation.
For the numerical approximation we rely on a well behaved ground state degeneracy as depicted later, so for all simulations fixed boundary conditions are chosen.
With periodic BC the ground state is highly degenerated, whereas with fixed BC the degeneracy of the ordered ground state is exactly $1$. 
For free BC the ground state is not the maximum ordered state of a triangular lattice.

It was proven that the number $\Omega(m,n)$ of different unimodular lattice triangulations on a grid $P_{m,n}$ grows exponentially with system size $mn$ \cite{Kaibel_2003}, hence, the entropy $S\propto\log\Omega$ is an extensive quantity.
To compare with the exact results and the bounds given in \cite{Kaibel_2003}, we use the capacity
\begin{equation}
\label{capacity} 
  C(m,n)\mathrel{:=} \frac {\log_2 \Omega(m,n)}{mn},
\end{equation}
which is basically the entropy of the system divided by the system size.

\section{Wang-Landau sampling}\label{sec:wang_landau}

Wang-Landau sampling  is a technique for numerically estimating the density of states (or the microcanonical entropy) of a system \cite{Wang_2001}, but can be generalised for arbitrary probability distributions. 
It is similar to the entropic sampling method of Lee \cite{Lee_1993}.
Both are flat histogram methods that try to establish a random walk hitting all energys with equal probability.

As most Monte Carlo algorithms the Wang-Landau sampling constructs a Markov chain proposing and accepting respectively rejecting elementary steps (for lattice triangulations these are the flips) so that the probability for finding a state in the chain follows a given probability distribution.
For flat histogram methods one samples according to $\Omega(E)^{-1}=e^{-S(E)}$, the inverse number of configurations with energy $E$.
Similar to the well-known Metropolis Monte Carlo scheme the acceptance probability of a flip from a configuration $\mu$ to a configuration $\nu$ has to be chosen as
\begin{equation}\label{eq:acceptance_wang_landau}
  P_{\mathrm{acc}}(\mu\to\nu)=\min\left(\frac {\Omega(E_\mu)}{\Omega(E_\nu)},1\right)=\min\left(e^{S(E_\mu)-S(E_\nu)}, 1\right)
\end{equation}
All possible steps from $\mu$ are proposed with equal probability, unflippable edges yield legitimate steps, but have $P_{\mathrm{acc}}=0$. 

However, the number of states $\Omega(E)$ with energy $E$ is not known a priori, so Wang-Landau sampling \cite{Wang_2001} proposes an iterative approximation scheme for the density of states:
Start with an estimation $\Omega_0(E)$ (e.g. $\Omega_0(E) = 1$ for all energies).
For each step the approximate number of states is updated at the visited energy according to $\Omega(E) \leftarrow \Omega(E)\cdot m_0$.
If the histogram $H(E)$ of visited energies is sufficiently flat, i.e. $\mathrm{min}\left\{H(E)\right\} > f\cdot \mathrm{avg}\left\{H(E)\right\}$ for a given $0 < f < 1$, the histogram $H(E)$ is reset and the modification factor $m$ is lessened according to $m_{i+1} = m_{i}^c$ with $0 < c < 1$ (typically $f \gtrapprox 0.8$ and $c \gtrapprox 0.8$ are chosen by experience).
The resulting $\Omega(E)$ is assumed to have converged if the modification factor becomes smaller than a predefined final value $m_f$.

Using Wang-Landau sampling for estimating the number of states $\Omega(E)$ with energy $E$ allows to numerically estimate the total number of states $\Omega = \sum_E \Omega(E)$.
However, the algorithm estimates the number of states $\Omega(E)$ only up to a multiplicative factor, which is not important for calculating expectation values.
To fix this multiplicative factor the degeneracy of a single energy level must be known.

\section{Approximate enumeration of  triangulations}

\subsection {Topological Energy}

\begin{figure}[ht]
  \centering
  \includegraphics[width=\columnwidth]{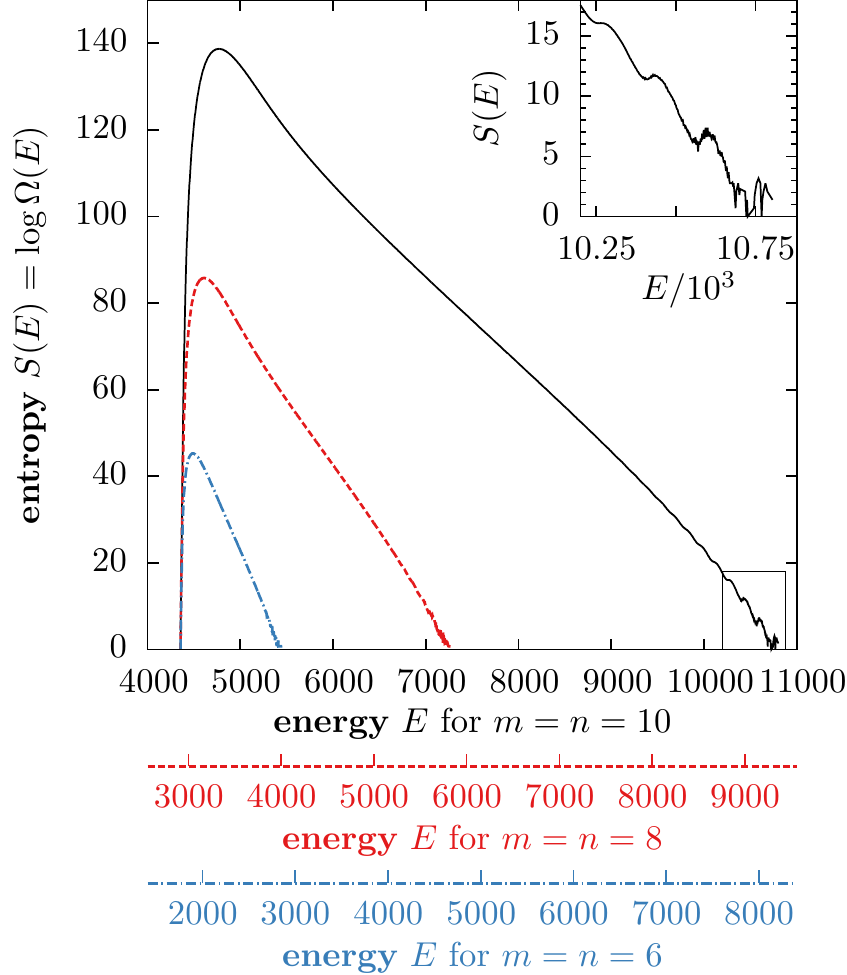}
  \caption{\label{fig:full_entropy}Microcanonical entropy $S(E) = \log \Omega(E)$ of lattice triangulations, calculated using Wang-Landau sampling with the vertex-degree-energy (\ref{eq:energy}): $6\times 6$ (blue, dash-dotted), $8\times 8$ (red, dashed) and $10\times 10$ (black, solid).}
\end{figure}

The Wang-Landau-Method is originally designed to estimate densities of states in terms of the energy.
To use it for counting arbitrary geometric configurations, one has to divide the configuration space into distinct classes by defining an energy functional as discrimination criterion.
In principle this definition is arbitrary for this purpose, however, the choices differ in computational efficiency. 
A natural choice should be discrete valued, illustrative and cheap to calculate.
Additionally the degeneracy of at least one energy class should be known exactly for normalisation purposes and it should be calculated solely from topological parameters, so that the method can be easily generalised to triangulations of general point sets.
The energy functional used here is the sum 
\begin{equation}
\label{eq:energy} 
  E = \sum_{\mbox{vertices } v} d_v^2
\end{equation} 
over the squared vertex degrees, which are the number of incident edges at a vertex.
The squaring is necessary as the sum over all vertex degrees is constant.
Similar energy functions were already used for calculating mixing times of Glauber dynamic on lattice triangulations \cite{Caputo_2013} and for calculating graph properties closed surface triangulations \cite{Aste_2012,Kownacki_2004}.
Using this energy functional the microcanonical entropy can be calculated as displayed in Fig.\,\ref{fig:full_entropy}.

\subsection {Limits on system size}\label{subsec:limits}

The entropy landscape shows a steep rise for small energies. 
Then, it reaches a maximum and declines smoothly towards high energies, as can be seen in Fig.\,\ref{fig:full_entropy}. 
For the dense configuration space regions around the peak the common Wang-Landau approach is perfectly unproblematic, but difficulties for large systems arise in the low and high energy regions, where the simulation can get stuck for different reasons.

In the low energy regions there are huge entropy differences between neighbouring states. 
The degeneracy of the first excited state $\Omega(E^\prime = E - E_{\mathrm{gs}} = 4)$ for $m\times n$ lattice triangulations can be calculated to be
\begin{equation}\label{eq:degeneracy_first_excited}
  \begin{split}
    &\Omega(E^\prime = 4) = \sum_{i = 1}^m \sum_{j = 1}^n (m - i + 1)(n - j + 1) + \\
    &+ \sum_{j = 1}^n\sum_{i = 1}^{m - j} (m - (i+j) + 1)(n - j + 1) + \\
    &+ \sum_{i = 1}^m\sum_{j = 1}^{n - i} (n - (i+j) + 1)(m - i + 1) \overset{m = n}{\approx} n^4/2
  \end{split}
\end{equation}
So for example in a $10\times 10$ system the number of first excited states is $\Omega(E^\prime = 4)=5665$ compared to $\Omega(E=0) = 1$ for the single ground state. 
Now, imagine a simulation step where the system is in ground state. 
While all edges are flippable, the corresponding acceptance probability of any edge flip is then $P_{acc}=1/5665$, the probability decreasing with system size $n\times n$ like $n^{-4}$. 
Certainly this limits the treatable system size severely -- not only by means of runtime, but also by exceeding the numerical precision of common floating point arithmetics.

In high energy regions the immobility is caused by a lack of connection between states in the same energy region, i.\,e.\ in general no short flip paths exist between states with similar energy. 
Furthermore, most edges in high energy states are unflippable. 
A typical immobile high energy state is depicted in Fig.\,\ref{fig:triang_examples}.
As for the ground state the algorithm can get stuck for long time in a high energy state due to high rejection rates caused by huge entropy differences.

One common approach to fix high rejection rates in low temperature Metropolis simulations is to use a rejection-free algorithm, known as ``the $N$-fold way'' or continuous time algorithm \cite {Bortz_1975}, which can be combined with the Wang-Landau method \cite {Schulz_2002}.
The basic idea of the $N$-fold way is to accept every step and to correct for the average number of steps a normal algorithm would perform before leaving the state.
The $N$-fold way was implemented for the lattice triangulations, but did not lead to improvements of the simulation times or the accessible system sizes.

\begin{figure}[ht]
  \centering
  \includegraphics[width=\columnwidth]{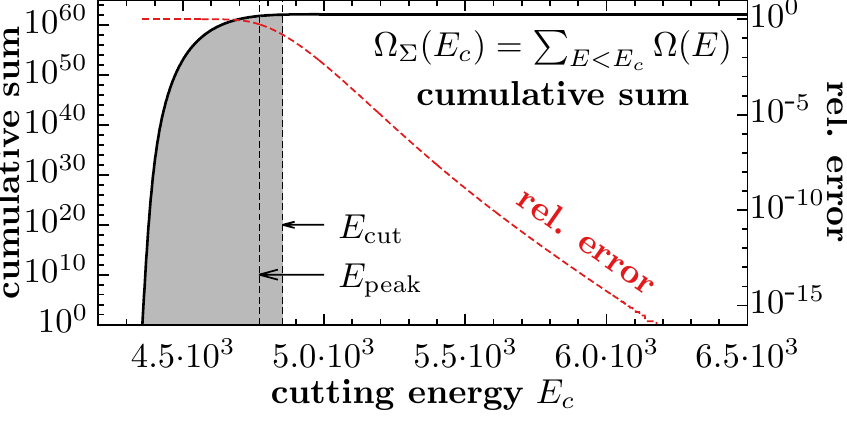}
  \caption{Cumulative sum $\Omega_\Sigma (E_c) = \sum_{E < E_c} \Omega(E)$ of the number of lattice triangulations $\Omega(E)$ in terms of the energy cutoff $E_c$ for a $10 \times 10$ lattice (black, solid line). The relative error $\Omega_\Sigma (E_c) / \Omega_\Sigma (\infty) - 1$ (red, dashed line) decreases rapidly for $E_c > E_{\mathrm{peak}}$, so that  only the grey part of the energy range can be used for calculating the number of lattice triangulations.}
  \label{fig:cumsum}
\end{figure}

The problem with high energy states can be overcome by defining an energy cut-off. 
Rejecting all steps beyond this energy leads to the correct estimate, as long as after each step -- even a rejected one -- the current state is correctly taken into account and added to the histogram of visits \cite{Schulz_2003}. 
This cut-off leads to a systematic underestimation of the total number of triangulations, yet, if the cut-off is chosen correctly the error is small, as can be seen from the cumulative sum $\Omega_\Sigma(E)=\sum_{E'<E} e^{S(E')}$ and the relative error $\epsilon_{\mathrm{cut}}(E) = (\Omega-\Omega_\Sigma(E)/\Omega)$ in Fig.\,\ref{fig:cumsum}. 
Unfortunately the low energies cannot be cut, as those are needed for normalisation. 
Furthermore, it is not known if ergodicity holds with a low energy cut-off. 
For high energy cut-offs ergodicity holds until a certain energy $E_{\mathrm{erg}} < E_{\mathrm{cut}}$, as the longest edge in a triangulations can always be shortened by a flip, and shorter edges correspond to lower energies.
The energy $E_{\mathrm{erg}}$ is big enough to leave the results unchanged.

\subsection{Energy cutoffs and initial estimates}
The question of finding a reasonable cut-off $\ecut{}$ remains. 
First the a-priori unknown energy range has to be estimated; while the energy $\emin{}$ of the ground state can be calculated, we approximate the maximal energy by constructing a star-shaped state of nearly maximum energy $\ehigh{}$ (cf. Fig.\,\ref{fig:triang_examples}). 
As can be seen in Fig.\,\ref{fig:full_entropy}, the entropy is peaked around an energy value of $\epeak{}$, that is easily accessible by simulating an unbiased random walk, i.\,e.\ a simulation where all proposed flips are accepted, and averaging over the sampled energies. 
These $3$ energies suffice to set a sensible energy cut-off
\begin{equation}\label{eq:energy_cutoff}
  \begin{split}
    \ecut{}&=\emin{}+\epsilon_{\mathrm{high}}(\ehigh{}-\emin{}) \textrm{ or }\\
    \ecut{}&=\emin{}+\epsilon_{\mathrm{peak}}(\epeak{}-\emin{})
  \end{split}
\end{equation}
where $\epsilon_{\mathrm{high}} < 1$ and $\epsilon_{\mathrm{peak}} > 1$ are constants that have to be fixed empirically.
Typical values can be $\epsilon_{\mathrm{high}} = 1/4$ and $\epsilon_{\mathrm{peak}} = 6/5$.

\begin{figure}[ht]
  \centering
  \includegraphics[width=\columnwidth]{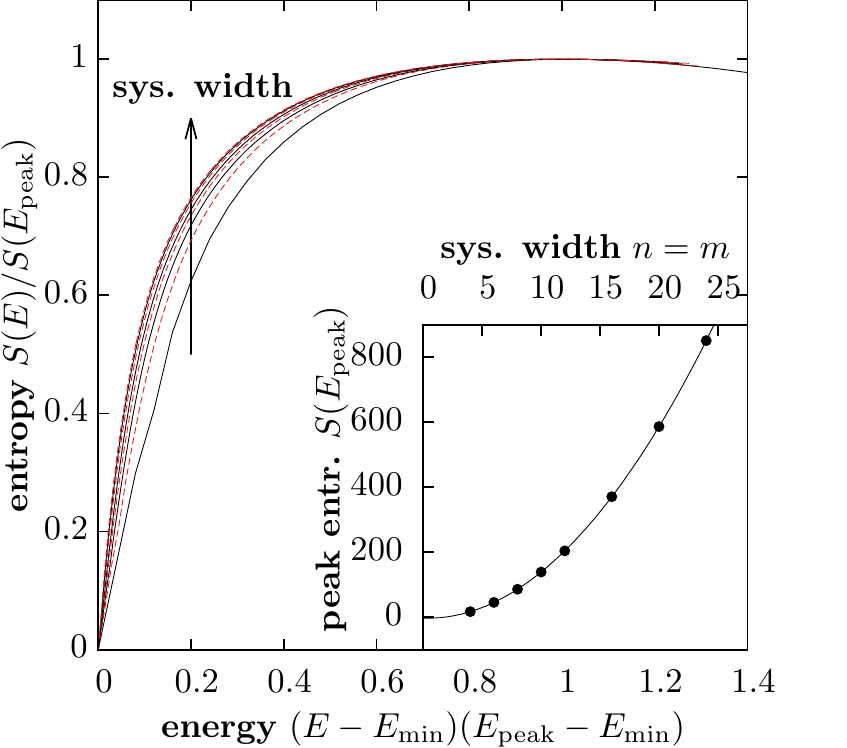}\\
  \includegraphics[width=\columnwidth]{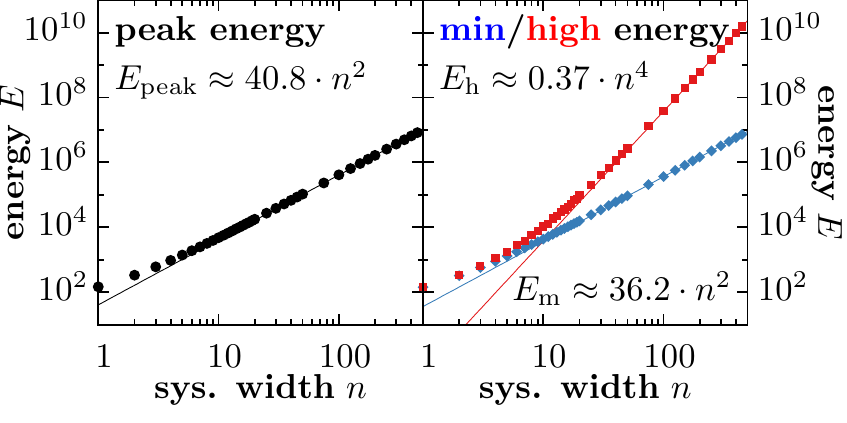}
  \caption{(a) Rescaled densities of state for quadratic lattices up to $n=25$. The densities converge, so that their extrapolation can be used as an initial estimation for simulations of larger lattices. (b) The peak and the minimal energy of triangulations as well as the peak energy entropy grow quadratically with the system width. The high energies, which are a lower bound for the maximal energy of  triangulations, grow with the system width to the power of 4.}
  \label{fig:rescaled_entropy}
\end{figure}

Extrapolating $S(\epeak{})$ from the Wang-Landau results for smaller systems an initial estimate for the entropy can be calculated for large systems. 
The precision of this extrapolation is not crucial.
Anyhow, the error is small, as can be seen in Fig.\,\ref{fig:rescaled_entropy}. 
The 4 parameters $\emin{}=6^2\cdot (n+1)^2$, $S(\emin{})=0$ (by definition of boundary conditions), $\epeak{}$ (measured or extrapolated) and $S(\epeak{})$ (extrapolated) characterise the entropy curve for quadratic lattices larger than $10\times 10$ sufficiently well (cf. Fig.\,\ref{fig:rescaled_entropy}). 

Using an initial estimate can speed up the relaxation process.
For $15\times 15$ triangulations the simulation with initial estimate extrapolated from $10 \times 10$ triangulations is by a factor of 3 faster than a simulation without one.
For larger systems the speedup is even more drastically.

\section{Results}

We use Wang-Landau sampling for an approximate enumeration of lattice triangulations to calculate the capacity for the different lattice sizes.
For all system sizes $5$ independent runs are performed. 
Beginning with modification factor $m_0=\exp \left(10^{-2}\right)$, it is reduced with the exponent$c=0.9$ whenever flatness $f=0.8$ is reached in the histogram of visits. 
The simulation is stopped, when modification factor $m_f=\exp\left( 10^{-12}\right)$ is reached.

\begin{table}[ht]
  \centering
  \begin{tabular}{|r|r@{.}l|}
    \hline
    $m$ & \multicolumn{2}{r|}{capacity $C$} \\
    \hline
    1 & ~~~0 & 00000 \\
    2 & 1 & 39657 \\
    3 & 1 & 66927 \\
    4 & 1 & 81445 \\
    5 & 1 & 90071 \\
    6 & 1 & 95728 \\
    7 & 1 & 99535 \\
    8 & 2 & 02433 \\
    \hline
  \end{tabular}
  \begin{tabular}{|r|r@{.}l|}
    \hline
    $m$ & \multicolumn{2}{r|}{capacity $C$} \\
    \hline
    9 & ~~~2 & 04615 \\
    10 & 2 & 06343 \\
    11 & 2 & 07745 \\
    12 & 2 & 08887 \\
    13 & 2 & 09819 \\
    14 & 2 & 10617 \\
    15 & 2 & 11281 \\
    16 & 2 & 11917 \\
    \hline
  \end{tabular}  \begin{tabular}{|r|r@{.}l|}
    \hline
    $m$ & \multicolumn{2}{r|}{capacity $C$} \\
    \hline
    17 & ~~~2 & 12374 \\
    18 & 2 & 12857 \\
    19 & 2 & 13263 \\
    20 & 2 & 13628 \\
    21 & 2 & 13858 \\
    22 & 2 & 14168 \\
    23 & 2 & 14352 \\
    24 & 2 & 14492 \\
    \hline
  \end{tabular}
  \caption{Capacity $C$  measured by the Wang Landau algorithm.}
  \label{tab:capacities}
\end{table}

\begin{figure}[ht]
  \centering
  \includegraphics[width=\columnwidth]{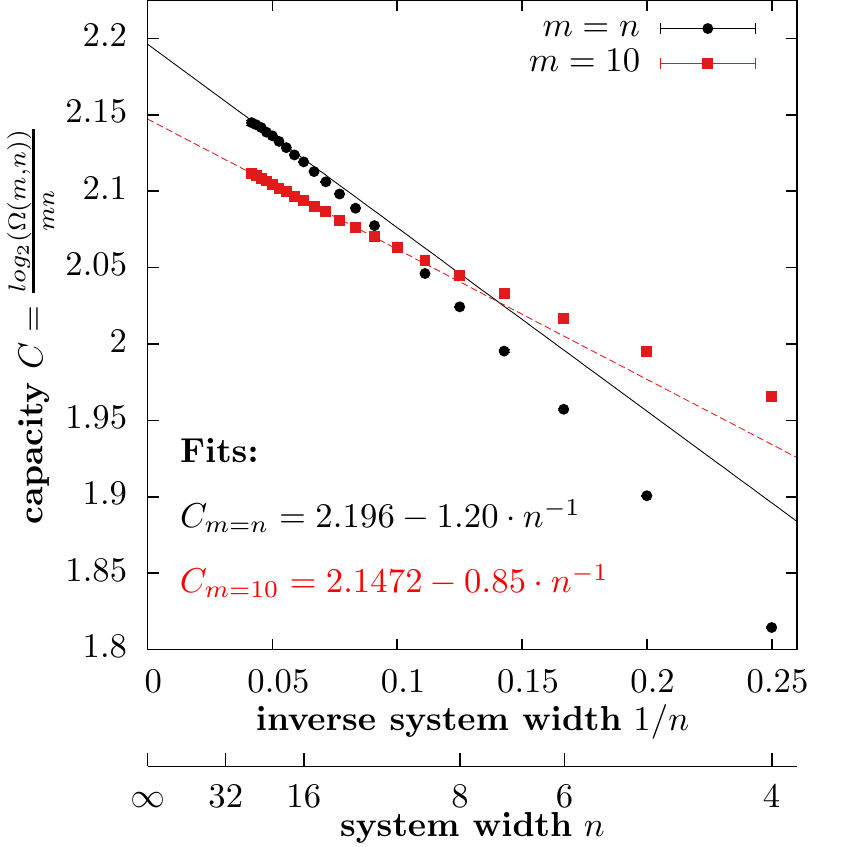}
  \caption{\label{fig:capacity} Measurements of the entropy density for systems up to size $24\times 24$.  
In the limit  $n \rightarrow \infty$ the extrapolated values are $C_{10} = 2.1472(4)$ for stripes of width $m$  and $C = 2.196(3)$ for quadratic lattices with $m=n$.}

\end{figure}

The validity of the method can be checked against the exact results of Kaibel and Ziegler \cite{Kaibel_2003} for small lattice sizes. 
The entity of interest is the capacity defined by Eq.\,\eqref{capacity} which is equivalent to the physical entropy density. 
As the entropies for different energies vary over large ranges, summation is done using multiprecision arithmetics from Python mpmath/gmp \cite{mpmath}.
In Fig.\,~\ref{fig:capacity} simulation data for narrow lattice stripes are compared to the exact results. 
All measurements are averaged over 5 independent runs, for large lattices a energy cutoff \eqref{eq:energy_cutoff} with the empirical $\epsilon_{\mathrm{high}} = 0.75$ is used.
For almost all considered system sizes the relative error of the simulation data is below $0.02$. 

\begin{figure}[ht]
  \centering
  \includegraphics[width=\columnwidth]{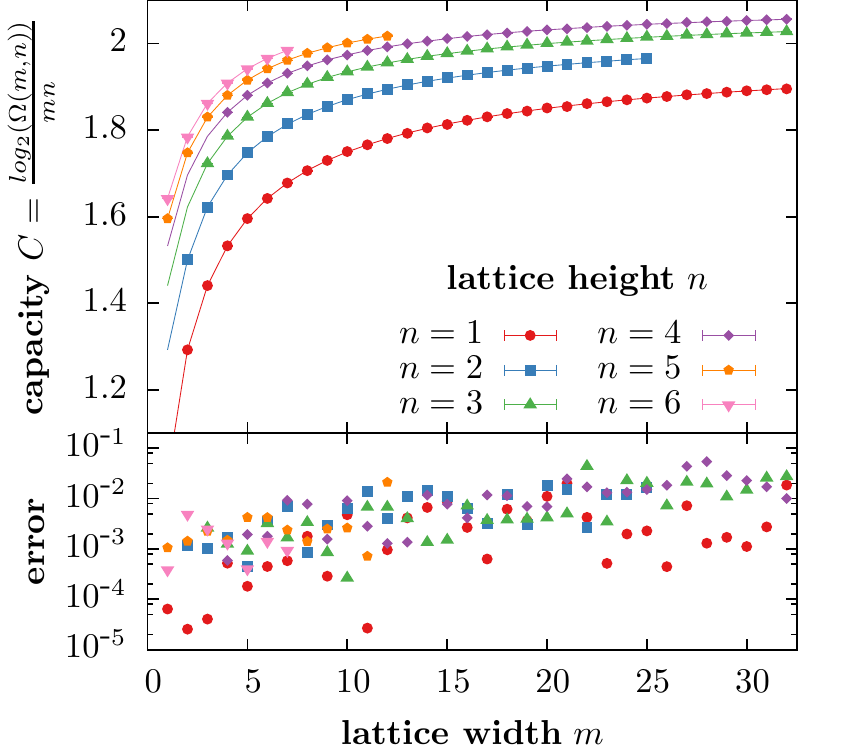}
  \caption{\label{fig:capacity_comparison} (a) Capacity $C(m,n) = (mn)^{-1} \log_2 \Omega(m,n)$ for triangulations of lattices with height $1 \leq n \leq 6$ (colour code) in terms of the lattice width $m$, calculated using Wang-Landau sampling. (b)  Relative error $C(m,n) / C_{\mathrm{exact}} - 1$ compared to the exact enumeration results of \cite{Kaibel_2003}.}
\end{figure}

The same is done for rectangular triangulations up to size $24 \times 10$ and quadratic triangulations up to size $24\times 24$, where the energy cutoff \eqref{eq:energy_cutoff} with the empirical $\epsilon_{\mathrm{peak}} = 1.2$ is used and an average over 3 independent runs was performed.
The initial entropy estimations $S_0(E)$ for systems of size $m=n>16$ are step-wise extrapolated from the relaxed result for smaller systems. 
Systems larger than including $20\times 20$ did not reach their final modification factor $m=10^{-12}$ in time. 
However, the results did not change any more during the last steps. 
This is an indication that saturation of error was already reached.

The capacities for the quadratic lattices are listed in Tab.\,\ref{tab:capacities}.
In Fig.\,\ref{fig:capacity} the capacity for the quadratic and rectangular triangulations is displayed.
Using a fit the limit of the capacity for infinite lattices and the asymptotic behaviour can be found to be
\begin{equation*}
  \begin{split}
    C_{m=10} &= (2.1472 \pm 0.0004) - (0.852 \pm 0.008)\cdot n^{-1} \\
    C_{m=n} &= (2.196 \pm 0.003) - (1.20 \pm 0.07)\cdot n^{-1}\;. 
  \end{split}
\end{equation*}

\section {Discussion}

It has been shown that approximate counting of lattice triangulations  is possible and feasible by using a Wang-Landau Monte-Carlo scheme. 
Our results for the capacity for lattice sizes below $24 \times 24$ improve the analytical bounds found in \cite{Kaibel_2003}.
As exact numbers are known for small systems, lattice triangulations provide a complement to the commonly used variants of the Ising model and other spin models.

One approach for optimising the measurement efficiency and possible access bigger lattices could be a different choice of the energy functional. 
Thereby, the challenge of high entropy differences between flip-connected states could be tackled as well. 
As a different approach, optimised probability weights other than the flat histogram probabilities could help to improve the sampling of neuralgic configuration space areas. 
Different algorithmic approaches like the transition matrix Monte Carlo algorithm \cite {Wang_2002} should be tested against the problem as well.

A generalisation of the estimation scheme could be interesting especially for mathematicians dealing with combinatorics. 
Scanning the energy landscape of a problem yields insight, it helps in winnowing dead ends from promising questions and generates a first estimate of what results to expect and quickly leads to interesting conjectures.
Furthermore it can be applied to interesting counting problems in mathematics, informatics and physics.


\bibliographystyle{eplbib}
\bibliography{literature_lattice_triangs}

\end{document}